\documentclass[10pt,letterpaper]{article}
\usepackage[top=0.85in,left=2in,footskip=0.75in,marginparwidth=2in]{geometry}

\usepackage[utf8]{inputenc}
\usepackage[toc,page]{appendix}

\usepackage{cite}

\usepackage{nameref,hyperref}

\usepackage[right]{lineno}

\usepackage{microtype}
\DisableLigatures[f]{encoding = *, family = * }

\raggedright
\usepackage{changepage}

\usepackage[aboveskip=1pt,labelfont=bf,labelsep=period,singlelinecheck=off]{caption}

\makeatletter
\renewcommand{\@biblabel}[1]{\quad#1.}
\makeatother

\usepackage{lastpage,fancyhdr,graphicx}
\usepackage{epstopdf}
\fancyheadoffset[L]{2.25in}
\fancyfootoffset[L]{2.25in}

\usepackage{color}

\definecolor{Gray}{gray}{.25}

\usepackage{graphicx}

\usepackage{sidecap}

\usepackage{wrapfig}
\usepackage[pscoord]{eso-pic}
\usepackage[fulladjust]{marginnote}
\reversemarginpar

\usepackage{graphicx} 
\usepackage{amssymb}
\usepackage{mathtools}
\usepackage{placeins}
\usepackage{gensymb}
\usepackage{graphicx}
\usepackage{subfig}
\usepackage{float}
\usepackage{bbm}

\begin{document}
\vspace*{0.35in}

\begin{adjustwidth}{-1.2in}{0in}

{\Huge
\textbf\newline{Collective Dynamics in Spiking Neural Networks Beyond Dale’s Principle}
}
\newline
\begin{flushleft}
{ 
Ross Ah-Weng\textsuperscript{1}, and
Hardik Rajpal\textsuperscript{2}
}
\end{flushleft}

\end{adjustwidth}
\begin{flushleft}
{
\small
\bigskip
\bf{1} Department of Computer Science, University College London; ross.ah-weng.25@ucl.ac.uk
\\
\bf{2} I-X Centre for AI in Science, Imperial College London; h.rajpal15@imperial.ac.uk
\\
}
\end{flushleft}

\section*{Abstract}
Dale's Principle has historically guided neuroscience research as a valuable rule of thumb, namely that all synapses on each neuron release the same set of neurotransmitters. Most existing Spiking Neuron Network models share this dichotomous assumption that neurons are either excitatory or inhibitory; however, recent experimental evidence points towards co-release mechanisms that violate this assumption. Here, we introduce a minimal model of "Bilingual" neurons violating Dale's principle that can exert both excitatory and inhibitory effects. We identify parameter regimes in which this architecture exhibits transitions between synchronous and asynchronous dynamics that differ quantitatively from those observed in a matched monolingual control architecture. We report distinct information-processing signatures both at the level of neurons and higher-order interactions between them near the phase transitions. These results suggest that the population of neurons violating Dale’s principle may provide an alternative mechanism for regulating large-scale oscillatory activity in neural circuits.

\section*{Introduction}
Spiking Neural Networks (SNNs) offer a biologically inspired approach to modelling brain dynamics, comprising a system of coupled non-linear ordinary differential equations. Each represents interacting neurons on a network that communicate via discrete-time events corresponding to the membrane potential hitting a threshold value and resetting to a lower value. The biological plausibility of SNNs makes them a powerful tool for studying complex collective neural phenomena, for example, synchronisation in modelling healthy brain function, such as beta and gamma rhythms \cite{FRIES2015220, sync, Aru005926}, but also pathological conditions such as epileptic seizures \cite{Jiruska2012-qq, MORMANN2003173}.

Moreover, SNNs have been instrumental in exploring the Critical Brain Hypothesis, which argues that the brain acts as a dynamical system that self-organises to the vicinity of a critical point between order and disorder \cite{OBYRNE2022820}. Signatures of criticality have been observed in vitro \cite{plenz}, and in vivo \cite{Hahn2010} neural activity. Near the vicinity of the critical point, the brain is thought to optimise information processing \cite{Shew55}, dynamic range \cite{kinouchi2006optimal} and promote healthy neural activity \cite{massobrio2015}. Order-disorder phase transitions, such as shifts from synchronous to asynchronous spiking regimes, have extensively been modelled using SNNs through network architecture \cite{ms17}, synaptic plasticity \cite{arc, stdp} and balance between excitatory and inhibitory neurons \cite{shewli}. Such an analysis is useful in identifying biological mechanisms that drive or regulate collective behaviour among neurons. This offers insights into broader cognitive phenomena, such as transitions between conscious and unconscious states \cite{fontenele19}.

The neurotransmitter co-release mechanism, in which multiple neurotransmitters are simultaneously released at the same vesicle, has been a topic of recent study \cite{Hnasko2011-az, Vaaga2014-qu, Tritsch2016-az, svensson19, Wallace2023-ln}. In particular, mounting evidence points to the existence of neurons that co-release Glutamate and Gamma-aminobutyric acid (GABA) \cite{Uchida2014, Yoo2016, Ntamati2016-wb, Noh2010-kt, Ceballos2024}, the former dubbing these "Bilingual Neurons". Barranca et al. argue that a single population of these neurons display a higher degree of balance than its two-population counterpart but exhibits a slower tracking of inputs \cite{Barranca2022-cr}. However, the dynamical properties of these systems remain unexplored, leaving open questions about how they differ from existing SNN models that follow Dale's principle.

This paper addresses this gap by modelling, in silico, the collective dynamics of a balanced population of Bilingual neurons. By tuning properties of the connectivity and base current, we observe novel transitions in their collective dynamics. Given the presence of a stable limit cycle, we find that the quenched fluctuations in the network may either couple or decouple the dynamics of neurons for a specific range of coupling strength. Furthermore, we observe that a sufficient level of disorder can drive the system out of equilibrium for progressively longer periods, even in a quiescent regime. 

We employ information-theoretic measures to characterise the spiking dynamics of individual neurons \cite{Strong1998-mo, luczak} and uncover pairwise and higher-order interactions \cite{Martignon1995-yf, Yu2011-bp}. These metrics allow us to distinguish between distinct dynamical regimes based on information processing, such as synchronous oscillations, often marked by high redundancy \cite{Narayanan2005-xc}, and asynchronous regimes. Using these measures, we characterise the emergent collective dynamics, associated information-processing structure, and the potential regulatory effects of mixed-sign synaptic architecture.

\section*{Materials and Methods}

We use a pulse-coupled network of $N$ Izhikevich neurons \cite{izh}, with dynamics described by the following system of differential equations:

\begin{equation}
    \begin{aligned}
        v'_i &= 0.04v_i^2 + 5v_i + 140 - u_i + I_i \\
        u'_i &= a(bv_i - u_i) \\
        v_i &\geq \theta \Rightarrow v \mapsto c, \; u_i \mapsto u_i + d
    \end{aligned}
\end{equation}

Here, $v_i, u_i$ denote the membrane potential and membrane recovery variable of neuron $i$, respectively, and time $t$ is scaled in milliseconds. These physical units follow the original parametrisation of the model, which is appropriately scaled to reproduce biologically realistic voltage dynamics. A spike is registered when $v_i$ reaches a threshold $\theta$, after which the reset map $v \mapsto c, \; u_i \mapsto u_i + d$ is instantaneously applied. We use the following standard parameters in \cite{izh} to model a homogeneous population of cortical neurons in the Regular Spiking regime, where $a = 0.02, \ b = 0.20, \ c = -65, \ d = 8, \theta = 30$.

The synaptic input is modelled as the sum of a base current $I_b$ and $\eta_i(t)$, whereby a presynaptic spike from neuron $i$ contributes an instantaneous pulse of amplitude $w_{ij}$ after a homogenous unit delay $\tau=1$: 
\begin{equation}
    I_i = I_b + \eta_i(t), \qquad \eta_i(t) \coloneqq \sum^N_{j \neq i}{\sum_{k}{w_{ij}\delta(t -t_k^j - \tau)}}
\end{equation}
where $t_k^j$ is the $k^{th}$ spike time of neuron $j$, and weights are sampled i.i.d about a Normal distribution symmetric about the origin, with a standard deviation $\frac{\sigma}{\sqrt{N-1}}$; this scaling factor ensures that the variance of the total synaptic input to each neuron is $O(1)$ as $N \to \infty$, consistent with classical mean-field scaling.
\begin{equation}
    w_{ij} \sim N(0, \frac{\sigma^2}{N-1})
\end{equation}

We thus introduce quenched disorder in the synapse distribution, where each neuron excites approximately half of its postsynaptic targets  ($w_{ij}>0$) and inhibits the complementary set of targets ($w_{ij} < 0$), in an all-to-all network. This models an idealised, minimal "Bilingual" architecture, with a symmetric Excitatory/Inhibitory ratio and mean in-degree of zero across the network.

We also consider the "monolingual" model as a control system, equivalent to a fully connected Pyramidal inter-neuronal Gamma (PING) architecture \cite{WHITTINGTON2000315}:

\begin{equation}
    S_i \sim B(1,0.5), \qquad
    \tilde{w}_{ij} \sim 
    \begin{cases}
    \vert w_{ij} \vert \quad \text{if} \ S_i = 0 \\
    -\vert w_{ij} \vert \quad \text{if} \ S_i = 1
    \end{cases}
\end{equation}
Since a zero-mean Gaussian variable is symmetric, multiplying $\lvert w_{ij} \rvert$ by an independent Rademacher variable yields a variable identically distributed to $w_{ij}$.

We numerically simulated the SNN model in discrete time, with $N=1000$, and fixed timestep $dt$ = 1 ms, using the fourth-order Runge-Kutta method. Each simulation was randomly initialised uniformly in the intervals $v_i\in[c, \theta], u_i\in[-15,0]$ for each neuron, to evaluate synchronisation properties from heterogeneous initial phases. These were run for 5000ms to allow the system to reach an assumed steady state, after which the firing events for the subsequent 10000ms were stored as binary spike trains. The Mean Firing Rate (MFR), $\bar{r}$, for the system was evaluated from times $t_0=5000$ to $T$ as follows:

\begin{equation}
    \bar{r} = \frac{1}{T-t_0}\int_{t_0}^T{r(t) \,dt}, \qquad r(t) = \frac{1}{N} \sum^N_{j=1}{\sum_{k}{{ \delta(t-t_k^j)}}}
\end{equation}

where $r(t)$ is the population-averaged instantaneous firing rate at time $t$.

We then performed a parameter sweep using $I_b \in[3,5]$ and $\sigma \in [0,100]$ as control parameters, chosen to span quiescent, synchronous and asynchronous regimes. This consisted of 15 repetitions of the model for each pair of parameters in the steady state. Here, $I_b$ is the base current and $\sigma$ controls the standard deviation of the system. We then evaluated the following information-theoretic signatures on the spike trains to quantify neuronal behaviour and higher-order interactions.

First, we analysed the Entropy Rate of each binary spike train, i.e. the limiting Shannon entropy $H$ of the $t^{th}$ state $X_t$ given all of the past states $X_{t-1}, X_{t-2}, \dots, X_1$:
\begin{equation}
    H'_{\mu X} = \lim_{t \rightarrow \infty}{\frac{1}{t}H(X_t|X_{t-1}, X_{t-2}, \dots, X_1)}
\end{equation}

We used the normalised Lempel-Ziv Complexity measure, $c(X,T)$, which is an efficient estimator of the Entropy Rate,   \cite{lz76}:

\begin{equation}
    \begin{aligned}
        H'_{\mu X} \approx   \lim_{T \rightarrow \infty}\frac{c(X, T) \log{T}}{T}  
    \end{aligned}
\end{equation}

The average entropy rate of the spike train is estimated by averaging local LZ76 over windows of size $T=3000$, which is sufficient due to fast convergence \cite{lz76mediano,improv}.

To further characterise the dynamics at the level of individual neurons, we computed the Active Information Storage (AIS) of each spike train. AIS quantifies how much information about the following observation $X_{t+1}$ can be found in the past $k$ state $X_t^{-k}$ \cite{AIS}.

\begin{equation}
    AIS(X) = \lim_{k \rightarrow \infty}{I(X_t^{-k} ; X_{t+1})}
\end{equation}

We use the Java Information Dynamics Toolkit (JIDT) \cite{jidt} to calculate the AIS over the observed window with embedding dimension $k=10$.

To evaluate pairwise interactions between neurons, we estimate the discrete Mutual Information (MI) between $N$ random pairs of binary spike trains. Since all neurons exhibit statistically similar dynamics on the fully connected network, average mutual information among $N$ random pairs provides a reliable estimate of average pairwise interdependency in the system.

\begin{equation}
    I(X_i; X_j) = H(X_i) + H(X_j) - H(X_i,X_j)
\end{equation}

To evaluate triplet interactions, we evaluate the O-information and S-information \cite{oinfo} between $N$ random triplets of spike trains, which respectively measure the balance between redundant and synergistic interactions, and the total magnitude of higher order interactions.
\begin{equation}
    \begin{aligned}
        \Omega(X^n) &= TC(X^n)-DTC(X^n) \\
        \Sigma(X^n) &= TC(X^n)+DTC(X^n)
    \end{aligned}
\end{equation}

where $TC(X^n)$ is the Total Correlation and $DTC(X^n)$ is the Dual Total Correlation among a system of $n$ variables $X^n$. All three measures of interdependence: MI, O-Info and S-Info were calculated on the observed window using the discrete estimators included in the JIDT package \cite{jidt}.

\section*{Results}
We present the results of the analysis on a 2D phase space, where the two dimensions of interest are the base current $I_b$ and the standard deviation of the weight distribution $\sigma$.

\begin{figure}[h]
    \centering
    \begin{minipage}{0.45\textwidth}
        \centering
        \includegraphics[width=\textwidth]{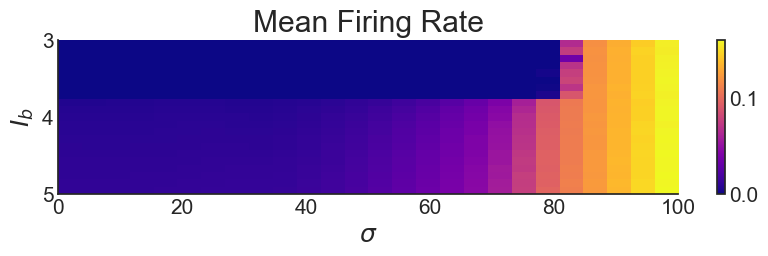}
    \end{minipage}
    \begin{minipage}{0.45\textwidth}
        \centering
        \includegraphics[width=\textwidth]{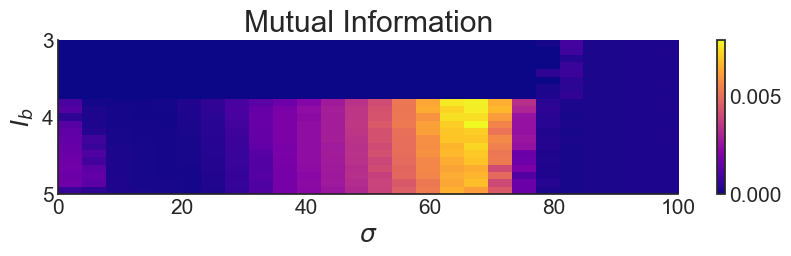}
    \end{minipage}

    \vspace{0.2cm} 

    \begin{minipage}{0.45\textwidth}
        \centering
        \includegraphics[width=\textwidth]{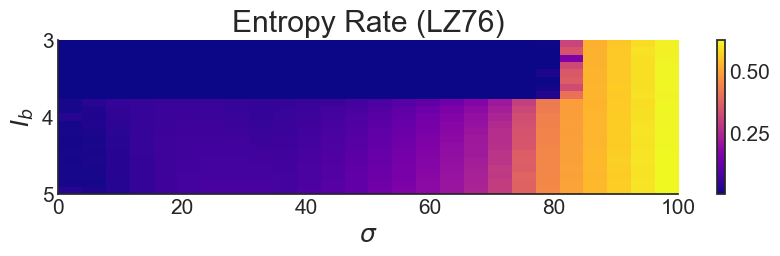}
    \end{minipage}
    \begin{minipage}{0.45\textwidth}
        \centering
        \includegraphics[width=\textwidth]{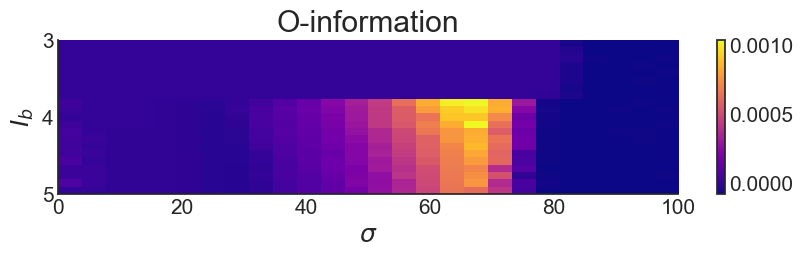}
    \end{minipage}

    \vspace{0.2cm}

    \begin{minipage}{0.45\textwidth}
        \centering
        \includegraphics[width=\textwidth]{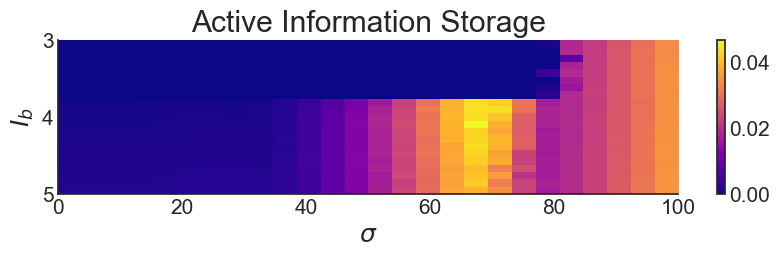}
    \end{minipage}
    \begin{minipage}{0.45\textwidth}
        \centering
        \includegraphics[width=\textwidth]{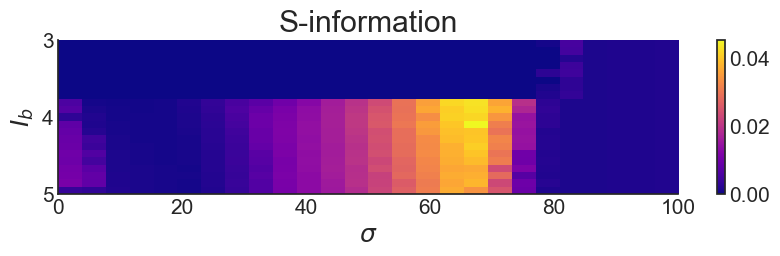}
    \end{minipage}

    \caption{\textit{Comparison of MFR and Information-Theoretic Measures in the Bilingual model.} Each heatmap illustrates their dependence on the base current $I_b$ (y-axis) and the standard deviation of the weight distribution $\sigma$ (x-axis), revealing phase transitions and shifts in information dynamics.}
    \label{bilingual}
\end{figure}

For sufficiently low variance $\sigma$, we observe a first-order transition in the MFR of both models from a quiescent regime to an active spiking regime [Figure \ref{bilingual}]. In the absence of network interactions, the Izhikevich neuron has a Subcritical Hopf bifurcation at $I_b = 3.7975$ at our chosen parameters [Appendix \ref{appendix:a}]. This is a generic mechanism in excitability transitions for type-II neurons \cite{izhbook}. In this case, a stable fixed point, representing inactivity, becomes unstable as the input current surpasses a critical threshold. This causes the neuron to jump to a large-amplitude stable limit cycle, formed from a fold limit cycle bifurcation for $I < I_b^c$ \cite{nesb}. A full bifurcation analysis of the coupled system is beyond the scope of the present study.

\begin{figure}[h]
    \centering
    \begin{minipage}{0.45\textwidth}
        \includegraphics[width=\textwidth]{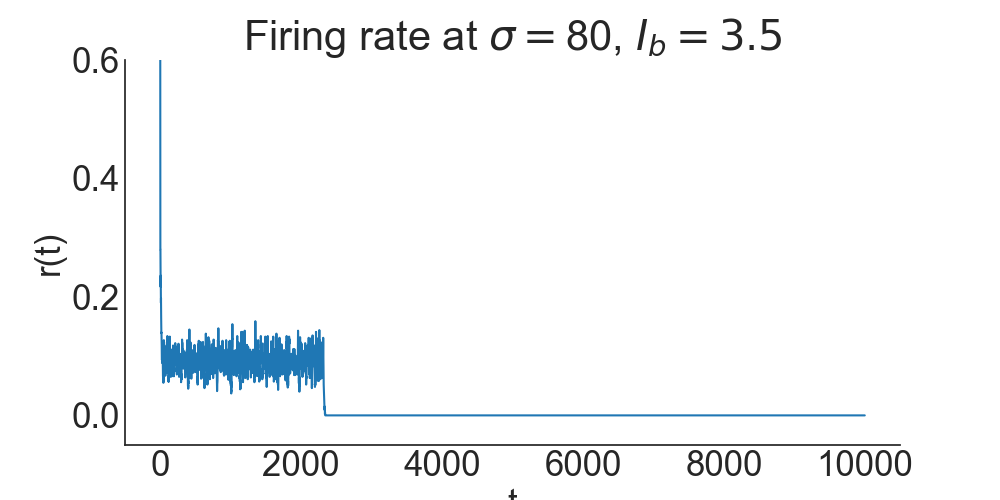} \\
        \includegraphics[width=\textwidth]{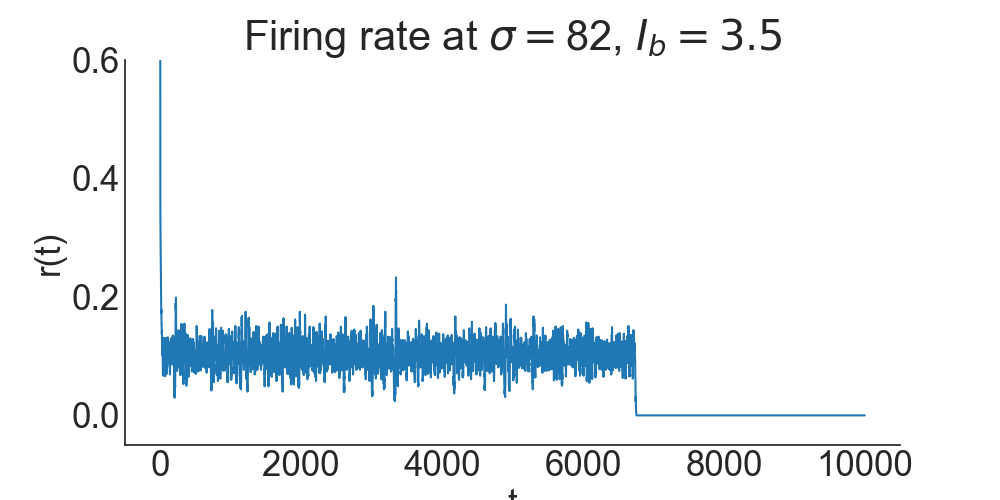} 
    \end{minipage}
    \hfill
    \begin{minipage}{0.5\textwidth}
        \centering
        \includegraphics[width=\textwidth]{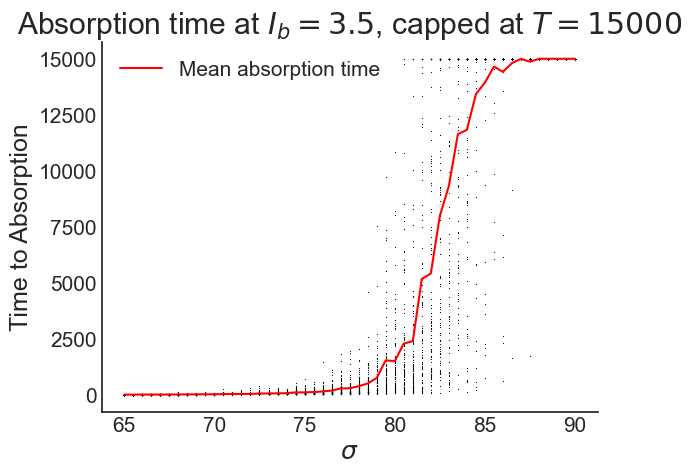}
    \end{minipage}
    \caption{\textit{Out of equilibrium behaviour in the Bilingual model}. The plot shows the mean duration of initial noise-driven activity against $\sigma$, capped at 15000ms. As $\sigma$ increases, the system transitions from quiescence to a state where the system increasingly persists out of equilibrium before absorption.}
    \label{fig:absorb}
\end{figure}

However in the Bilingual model, we observe a sharp increase in mean firing rate (MFR) as $\sigma$ increases, below $I_b^c$. This is corroborated by a jump in LZ76 and AIS, indicating an increase in signal complexity and predictability, corresponding to neuronal bursting behaviour. However, the higher-order information-theoretic signatures, MI, O-information and S-information drop to close to $0$, indicating an asynchronous regime. This transition coincides with a sudden increase in the duration for which initial excitations persist before the system settles into its absorbing quiescent state (Figure \ref{fig:absorb}); for $I_b < I_b^c$, the uncoupled neuron possesses a stable fixed point. As $\sigma$ increases, the quenched disorder in the synaptic weights generates stronger fluctuation-driven excursions away from the fixed point. In simulations, this manifests as increasingly long-lived transient activity before eventual quiescence.

\begin{figure}[h]
    \centering
    \includegraphics[width=0.55\textwidth]{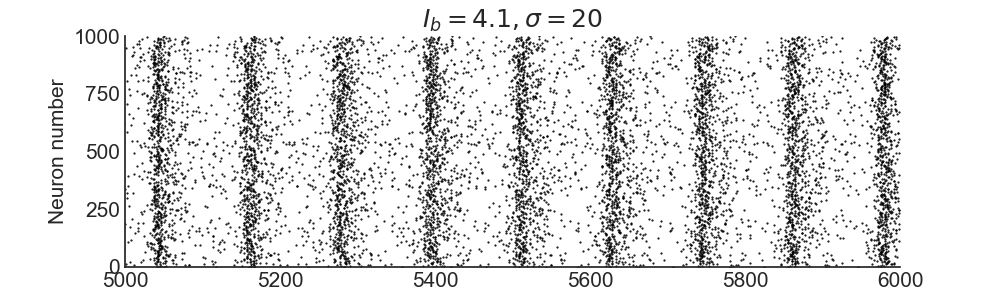}
    \includegraphics[width=0.55\textwidth]{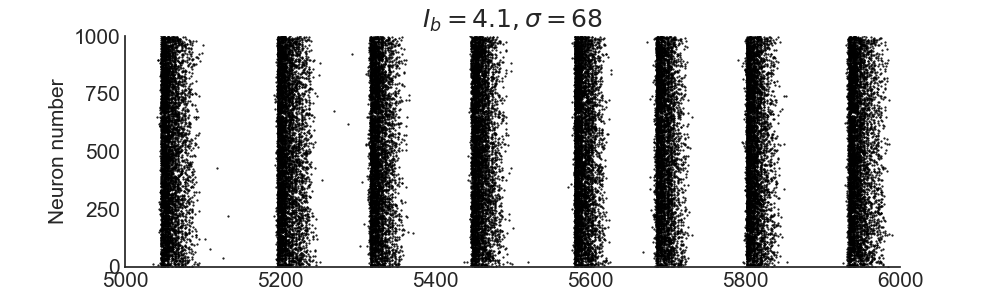}
    \includegraphics[width=0.55\textwidth]{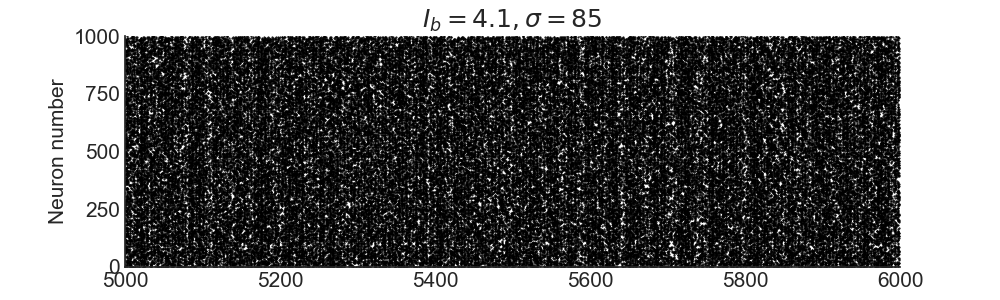}

    \caption{\textit{Raster plots of neuronal firing in the Bilingual model for weak-coupling, synchronous, and asynchronous regimes at $I_b = 4.1$.} In these regimes, we respectively observe Regular Spiking, Chattering and Stuttering behaviours of the individual Izhikevich neurons \cite{izhbook}.}
    \label{raster}
\end{figure}

In the Bilingual model, for $I_b > I_b^c$, we observe an increase in MFR and LZ76 with increasing $\sigma$ as seen in Figure \ref{fig:Entropy, AIS} - the higher the coupling strength, the more firing activity. For low $\sigma$, neurons are weakly coupled and, hence, oscillate at their natural frequency out of phase - reflected by low higher-order information signatures. However, as $\sigma$ increases, the system synchronises, and we observe clear peaks in AIS, MI, O-info and S-info at $\sigma \approx 68$, indicating the presence of a synchronous, oscillatory regime dominated by redundancy (indicated by the positive value of O-information). After this, for high $\sigma$, noise-driven fluctuations dominate collective oscillations, and the system transitions to an asynchronous regime as before, seen by the sustained AIS and drop in higher-order interactions. The three distinct behaviours for the populations of the neurons can be seen in Figure \ref{raster}. 

\begin{figure}[h]
    \centering
    \begin{minipage}{0.45\textwidth}
        \centering
        \includegraphics[width=\textwidth]{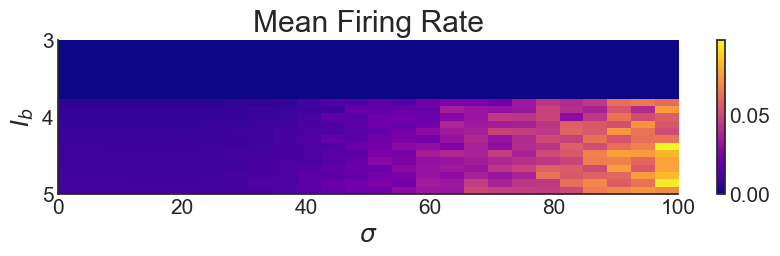}
    \end{minipage}
    \begin{minipage}{0.45\textwidth}
        \centering
        \includegraphics[width=\textwidth]{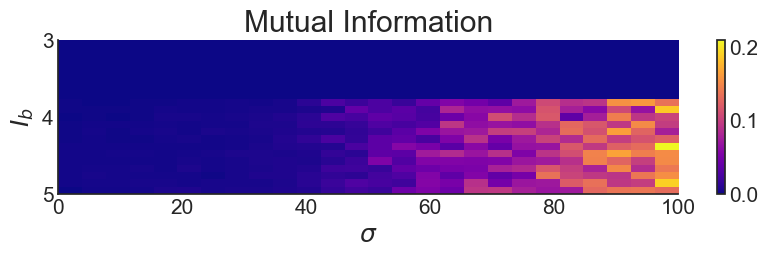}
    \end{minipage}

    \vspace{0.2cm} 

    \begin{minipage}{0.45\textwidth}
        \centering
        \includegraphics[width=\textwidth]{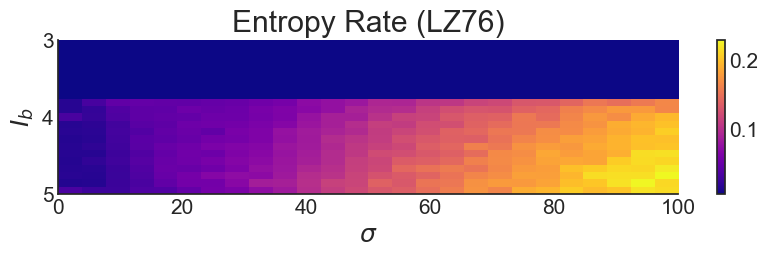}
    \end{minipage}
    \begin{minipage}{0.45\textwidth}
        \centering
        \includegraphics[width=\textwidth]{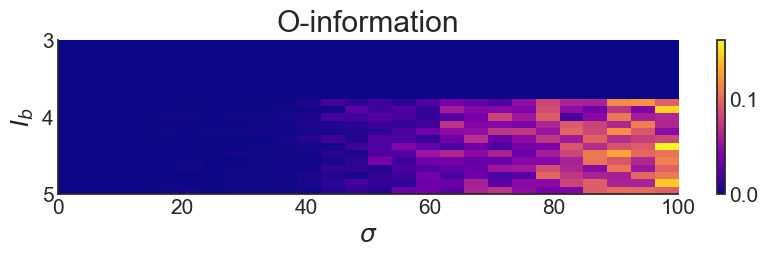}
    \end{minipage}

    \vspace{0.2cm}

    \begin{minipage}{0.45\textwidth}
        \centering
        \includegraphics[width=\textwidth]{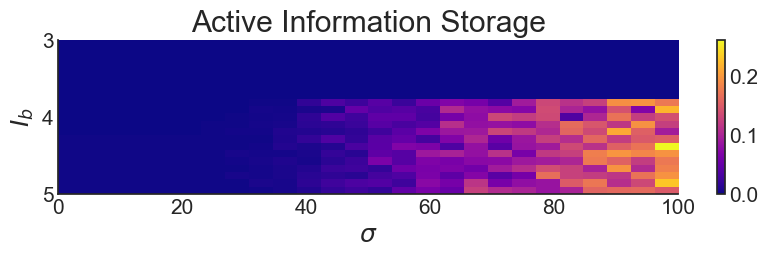}
    \end{minipage}
    \begin{minipage}{0.45\textwidth}
        \centering
        \includegraphics[width=\textwidth]{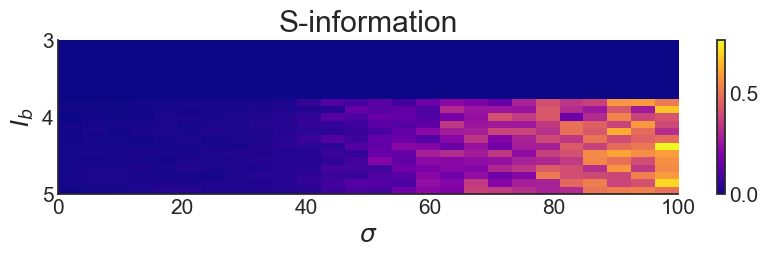}
    \end{minipage}

    \caption{\textit{Comparison of MFR and Information-Theoretic Measures in the Monolingual model.} From the heatmaps, we observe a simpler relationship between our control parameters $I_b$ (y-axis) and $\sigma$ (x-axis).}
    \label{monolingual}
\end{figure}

However, in the monolingual model, we observe a simple synchronisation transition with increasing $\sigma$ as all information-theoretic measures monotonically increase, as seen in Figure \ref{monolingual}. This aligns with classical Excitatory-Inhibitory PING architectures, and these results align with the existing literature on oscillations in these models \cite{WHITTINGTON2000315, nguyen21, Nandi2024-zr}. Moreover, the magnitude of the entropy rate is, on average, less than that of the Bilingual model, and the magnitude of MI and S-information is, on average, significantly higher than that of the Bilingual model. This suggests that the monolingual model exhibits more robust and synchronised oscillations, with stronger correlations between neuronal units marked by high redundancy in the system. However, information measures do not peak near the transition as seen in the Bilingual model and other complex systems with critical phase transitions~\cite{lizier2011information, marinazzo2019synergy, rajpal2025information}. 

\section*{Discussion}
Our findings demonstrate that Bilingual neurons, capable of co-releasing Glutamate and GABA, exhibit distinct dynamical phase transitions compared to traditional Monolingual excitatory-inhibitory (E-I) systems. Unlike classical SNN models in which phase transitions are attributed to spectral properties of the adjacency matrix \cite{shewli}, the empirical spectral distributions of the two architectures considered here appear to obey the same circular law and evolve smoothly with the coupling variance. These results suggest that the observed phase transitions are intrinsically dynamical or noise driven in mechanism. Notably, this minimal formulation does not require synaptic plasticity or neuronal heterogeneity. Despite homogeneous connectivity and a mean neuronal input of zero - conditions under which classical mean-field theory often fails to capture fluctuation-driven activity - the model remains structurally simple and potentially analytically tractable. This aligns with recent advances in modelling heterogeneous Izhikevich networks incorporating adaptation and delays \cite{Chen2022-jt, chen2024}.

In the presence of a limit cycle, by increasing the connection strength, we observe that this initially promotes synchronisation, but excessive strengthening destabilises coherent oscillations in the Bilingual architecture. This behaviour resembles desynchronisation mechanisms relevant to pathological oscillations in epileptic behaviour. However, we emphasise that our model is highly idealised and does not incorporate biophysical seizure dynamics; therefore, such interpretations should be regarded as speculative. Unlike classical E-I systems, where inhibition is typically imposed as a corrective mechanism after excitatory activity has already escalated, Bilingual neurons may allow for pre-emptive regulation of excitatory bursts in a temporally precise manner. This aligns with work on models of presynaptic inhibition \cite{naumann20}, as well as experimental studies \cite{gemin, Chiu2013-gq}. Such dynamics may optimise the trade-off between network stability and computational flexibility, potentially expanding the system’s information-processing capacity.

While our all-to-all model serves as a foundational abstraction, future work should impose biologically realistic modelling assumptions, such as imbalanced Excitatory-Inhibitory weights, alternative network topologies or adaptation of Spike-Timing-Dependent-Plasticity (STDP) like mechanisms for Bilingual neurons. The role of plasticity is crucial as it allows the synaptic weights to evolve over the course of dynamics; however, a proper plasticity mechanism for Bilingual neurons is yet to be formulated.

Mounting experimental evidence suggests that co-release, whilst only representing a small proportion of synaptic vesicles, nevertheless coexists alongside monolingual systems. This raises further questions about their integration and role in cortical circuits - addressing these will require combining our minimal model with data-driven connectomics and in vivo recordings of co-transmission. Our results demonstrate that relaxing Dale’s principle in a minimal spiking network model produces qualitatively distinct dynamical regimes. Whether such mechanisms operate in biological circuits remains an open question, requiring targeted experimental and modelling studies to understand neurotransmitter co-release and its implications for cognitive processes.


\subsection*{Author Contributions}
Conceptualisation, R.A. and H.R.; methodology, R.A. and H.R.; software, R.A.; validation, R.A. and  H.R.; formal analysis, R.A.; writing -- original draft preparation, R.A.; writing -- review and editing, H.R.; visualisation, R.A.; supervision, H.R. Both authors have read and agreed to the published version of the manuscript.

\subsection*{Funding}
R.A. was supported by the Department of Mathematics, Imperial College's Undergraduate Research Opportunity Programme (UROP), and is currently supported by the EPSRC Centre for Doctoral Training in Collaborative Computational Modelling at the Interface (Grant No. EP/Y034767/1). H.R. is supported by the Statistical Physics of Cognition project funded by the EPSRC (Grant No. EP/W024020/1) and by the Eric and Wendy Schmidt AI in Science Postdoctoral Fellowship.

\subsection*{Acknowledgments}
We thank Prof. Henrik Jeldtoft Jensen and Dr Pedro Mediano for their insightful comments. We also thank the Research Computing Services of Imperial College for providing access to the High Performance Computing cluster used for this study.

\subsection*{Abbreviations}{
The following abbreviations are used in this manuscript:\\

\noindent 
\begin{tabular}{c c}
SNN & Spiking Neural Network\\
GABA & Gamma-aminobutyric Acid\\
MFR & Mean Firing Rate\\
AIS & Active Information Storage\\
MI & Mutual Information\\
STDP & Spike-Timing-Dependent-Plasticity
\end{tabular}
}

\appendix
\appendixpage
\section{}
\label{appendix:a}
Assume $\eta_i(t)=0$. Finding the Jacobian for the system:

\begin{equation}
    J(v,u) =
    \begin{pmatrix}
        \frac{\partial v'}{\partial v} & \frac{\partial v'}{\partial u} \\
        \frac{\partial u'}{\partial v} & \frac{\partial u'}{\partial u}
    \end{pmatrix}
     =
    \begin{pmatrix}
        0.08v + 5& -1 \\
        ab & -a
    \end{pmatrix}
\end{equation}

Checking the conditions for Hopf Bifurcation using parameters $a=0.02, b=0.2$:

\begin{equation}
    Tr[J(v,u)] = 0 \iff v^* = \frac{a-5}{0.08} = -62.25
\end{equation}

\begin{equation}
    Det[J(v^*,u)] > 0 \iff b > a
\end{equation}

which are satisfied. Setting $v', u'$ to zero yields equation for the fixed point:

\begin{equation}
    0.04{v^*}^2 + (5-b)v^*+140+I_b = 0
\end{equation}

Substituting $v^*$ yields $I_b^c=3.7975$. Further work may be done to find the sign of the first Lyapunov exponent \cite{izhbook}:

\begin{equation}
    \text{sign}[l_1] = \text{sign} \left[F''' + \frac{(F'')^2}{b-a}\right] = \text{sign} \left[
    \frac{(0.08)^2}{(b-a)}
    \right] = +1
\end{equation}

where $F(v) = 0.04v^2 + 5v + 140 + I_b]$. This indicates the presence of a Subcritical Hopf Bifurcation.

\section{}
\begin{figure}[h]
    \centering
    \begin{minipage}{0.45\textwidth}
        \includegraphics[width=\textwidth]{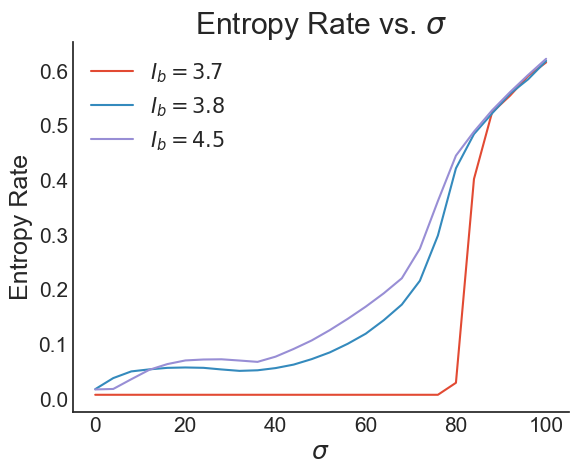} \\
        \includegraphics[width=\textwidth]{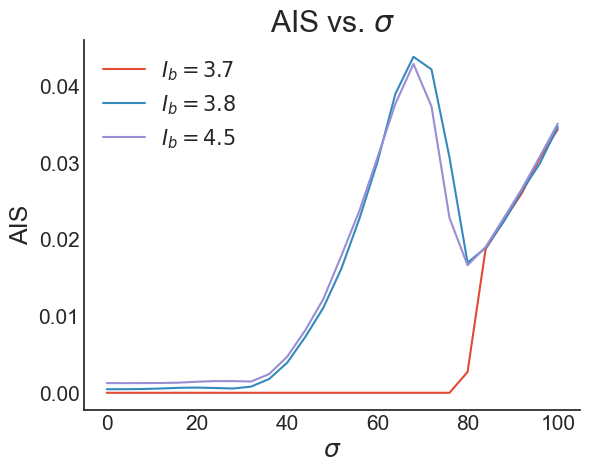} 
    \end{minipage}
    \hfill
    \begin{minipage}{0.45\textwidth}
        \centering
        \includegraphics[width=\textwidth]{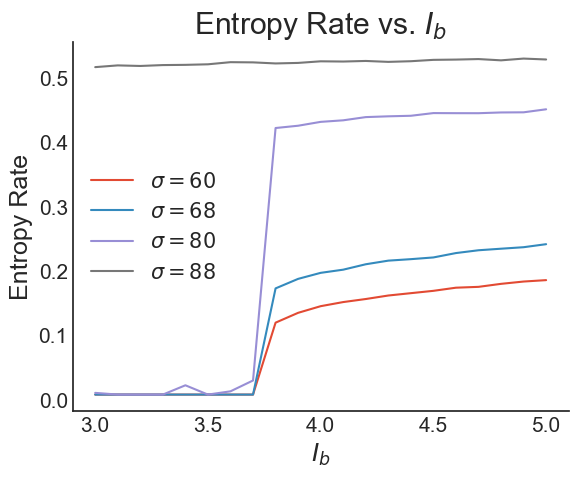} \\
        \includegraphics[width=\textwidth]{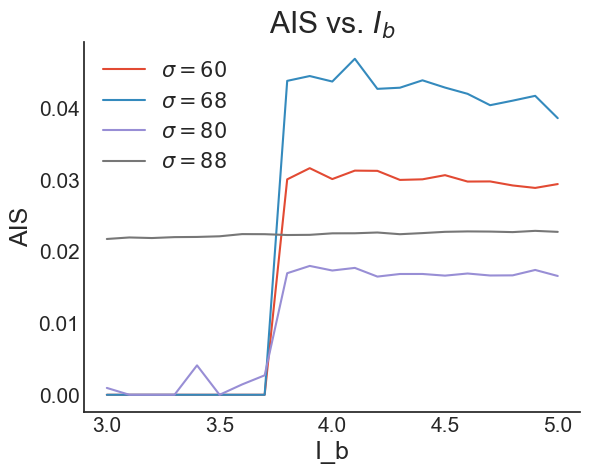} 
    \end{minipage}
    \caption{1D sections of Entropy Rate and AIS phase plots of the Bilingual model. }
    \label{fig:Entropy, AIS}
\end{figure}

For fixed $I_b < I_b^c$, we observe a transition in both Entropy Rate and AIS with respect to $\sigma$ characterised by a sharp, discontinuous increase. In contrast, for $I_b > I_b^c$, we observe a smoother increase in Entropy Rate in $\sigma$, eventually converging to the same value as observed for $I_b=3.7$. However, AIS detects a peak in synchrony at $\sigma \approx 68$, before dipping down again to align with the AIS for $I_b = 3.7$. This implies that for sufficiently high $\sigma$, asynchronous activity becomes dominant, overriding the influence of the base current.

Now fixing $\sigma$ and varying $I_b$, we reach the same conclusion. For low $\sigma$, the discontinuous jumps in Entropy Rate and AIS correspond to the bifurcation at $I_b^c$, marking a transition between inactive and active states. However, at high $\sigma$, Entropy Rate and AIS stabilise, indicating that noise-driven activity dominates over the bifurcation-driven transition. This implies that beyond a certain noise threshold, the system enters a regime where intrinsic dynamics are overshadowed by stochastic fluctuations, leading to a loss of structured information flow.

\nolinenumbers

\newpage

\begin{adjustwidth}{-1.45in}{0in}

\bibliography{bibliography}

\bibliographystyle{abbrv}

\end{adjustwidth}

\end{document}